\documentclass[%
aps,%
prb,%
twocolumn,%
showpacs,%
a4paper]%
{revtex4}
\usepackage{graphicx}
\usepackage[germanpar]{anysize}%
\usepackage{bm}
\usepackage{bbm}
\usepackage{stmaryrd}
\usepackage{color}%
\usepackage{amsmath,amssymb}

\marginsize{2cm}{1cm}{0.5cm}{2cm}

\newcommand{\tmfloatsmall}[2]{
\begin{figure}
#1
\caption{#2}
\end{figure}}

\newcommand{\emdash}{---}

\begin{document}

\title{Hofstadter butterflies of bilayer graphene}

\author{Norbert Nemec}
\author{Gianaurelio Cuniberti}
\affiliation{Institute for Theoretical Physics,
 University of Regensburg,
 D-93040 Regensburg, Germany}

\date{\today}

\begin{abstract}
  We calculate the electronic spectrum of bilayer graphene in perpendicular
  magnetic fields nonperturbatively. To accomodate arbitrary displacements
  between the two layers, we apply a periodic gauge based on singular flux
  vortices of phase $2 \pi$. The resulting Hofstadter-like butterfly plots
  show a reduced symmetry, depending on the relative position of the two
  layers against each other. The split of the zero-energy relativistic Landau
  level differs by one order of magnitude between Bernal and non-Bernal
  stacking.
\end{abstract}

\pacs{
73.22.-f,
71.15.Dx,
71.70.Di,
81.05.Uw
}

\maketitle


After the theoretical prediction of the peculiar electronic properties of
graphene in 1947 by Wallace{\cite{wallace-tbtog1947}} and the subsequent
studies of its magnetic spectrum,{\cite{mcclure-dog1956,zheng-hcoatgs2002}} it
took half a century until single layers of graphene could be isolated in
experiment{\cite{novoselov-tac2005}} and the novel mesoscopic properties of
these two-dimensional (2D) Dirac-like electronic systems, e.g., their
anomalous quantum Hall effect, could be
measured.{\cite{zhang-eootqheabpig2005,novoselov-tgomdfig2005,zhang-lsigihmf2006}}
Inspired by this experimental success, graphene has become the focus of
numerous theoretical
works.{\cite{gusynin-uiqheig2005,kane-qsheig2005,peres-epodtc2006,guinea-esalligs2006,hasegawa-qheattnig2006}}
For bilayers of graphene, an additional degeneracy of the Landau levels and a
Berry phase of $2 \pi$ were predicted to lead to an anomalous quantum Hall
effect, different from either the regular massive electrons or the special
Dirac-type electrons of single-layer graphene,{\cite{mccann-ldaqheiagb2006}}
which was confirmed in experiment shortly
afterwards{\cite{novoselov-uqheabpo2ibg2006}} and used for the
characterization of bilayer samples.{\cite{ohta-ctesobg2006}}

The low-energy electronic structure of a single layer of graphene is well
described by a linearization near the corner points of the hexagonal Brillouin
zone ($K$ points), resulting in an effective Hamiltonian formally equivalent
to that of massless Dirac particles in two
dimensions.{\cite{divincenzo-setfisigic1984}} A related Hamiltonian can be
constructed featuring a supersymmetric structure which can be exploited to
derive the electronic spectrum in the presence of an external magnetic
field.{\cite{ezawa-sauqheig2006}} The level at zero energy, characteristic for
any supersymmetric system, maps directly to a special half-filled Landau level
fixed at the Fermi energy $E_{\mathrm{F}}$, henceforth called the
\textit{supersymmetric Landau level} (SUSYLL).

\tmfloatsmall{\resizebox{\columnwidth}{!}{\includegraphics{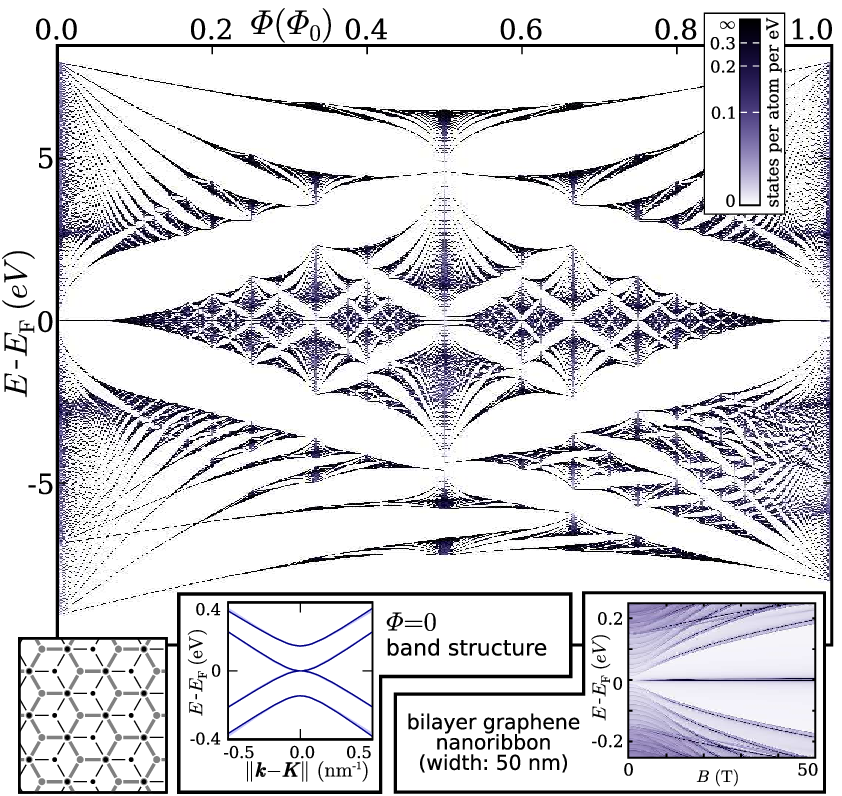}}}{\label{fig:butterfly-bernal}(Color
online) Hofstadter butterfly of a bilayer graphene in the Bernal stacking
configuration. The band structure at zero magnetic field is rotationally
symmetric in good approximation for an area around the $K$ point and shows a
split into four massive bands, with the two middle ones touching at
$E_{\mathrm{F}}$. The density of states (DOS) of a finite-width ribbon [a pair
of (200,0) zigzag ribbons] in the same configuration shows the SUSYLL emerging
at finite magnetic field. The split of the SUSYLL (discussed below) is not
visible due to the limited resolution of the plot.}

In this Rapid Communication, we use the nonperturbative method pio\-neer\-ed in
1933 by Peierls{\cite{peierls-ztddvl1933}} for the implementation of a
magnetic field in a model, which led Hofstadter, in 1976, to the discovery of
the fractal spectrum of 2D lattice electrons in a magnetic
field.{\cite{hofstadter-elawfobeiraimf1976}} Since its discovery, various
aspects of the so-called \textit{Hofstadter butterfly} have been
studied,{\cite{albrecht-eohfesitqhc2001,analytis-llmoathbifs2004}}
particularly in relation to graphenelike honeycomb
structures.{\cite{rammal-llsobeiahl1985,hasegawa-qheattnig2006,nemec-hbocnpotms2006}}
Featuring a large variety of topologies, all these systems have in common that
the atoms inside the unit cell are located on discrete coordinates. All closed
loops have commensurate areas, and the atomic network is regular enough that
the magnetic phases of all links can be determined individually without the
need of a continuously defined gauge field. For bilayer graphene, such a
direct scheme for implementing a magnetic field is possible only for highly
symmetric configurations like Bernal
stacking.{\cite{bernal-tsog1924,mccann-ldaqheiagb2006}} To handle more general
configurations, such as continuous displacements between the layers, it is in
general unavoidable to choose a continuously defined gauge that fixes the
phase for arbitrarily placed atoms. The difficulty that arises can be seen
immediately: For any gauge field that is periodic in two dimensions, the
magnetic phase of a closed loop around a single unit cell must cancel out
exactly, corresponding to a vanishing total magnetic flux. Conversely, this
means that any gauge field that results in a nonzero homogeneous magnetic
field will invariably break the periodicity of the underlying system.

A possible way to bypass this problem is based on defining a {\emph{magnetic
flux vortex}}, here oriented in the $z$~direction and located in $\left( x_0,
y_0 \right)$, as{\cite{trellakis-nsfbeicmf2003,cai-aiciaumfups2004}}
\begin{eqnarray*}
  \boldsymbol{B} \left( x, y, z \right) & = & \Phi_0 \delta \left( x - x_0
  \right) \delta \left( y - y_0 \right) \boldsymbol{e}_z,
\end{eqnarray*}
where $\Phi_0 = h / e$ is the flux quantum. Physically, such a vortex is
equivalent to a vanishing magnetic field, since it leaves the phase of any
possible closed path unchanged modulo $2 \pi$. One possible gauge field
resulting in such a single flux vortex can be written as
\begin{eqnarray*}
  \boldsymbol{A} \left( \boldsymbol{r} \right) & = & \frac{\Phi_0 \left(
  \boldsymbol{e}_z \times \boldsymbol{r} \right)}{2 \pi \left| \boldsymbol{e}_z
  \times \boldsymbol{r} \right|^2} .
\end{eqnarray*}
Finding a periodic gauge follows straightforwardly. To the homogeneous
magnetic field, we add a periodic array of flux vortices with a density such
that the average magnetic field is exactly zero. For the resulting field,
which is physically equivalent to the original, it is now possible to find a
gauge field with the same periodicity as the array of vortices. If the
underlying system is periodic and the array of flux vortices has commensurate
periodicity, there exists a supercell where the magnetic Hamiltonian is
periodic. One possible periodic gauge that is especially advantageous for
numerical implementation consists in a two-dimensional periodic system with
lattice vectors $\boldsymbol{a}_x$ and $\boldsymbol{a}_y$. The reciprocal lattice
vectors (scaled by $2 \pi$) are $\tilde{\boldsymbol{a}}_i$ such that
$\boldsymbol{a}_i \cdot \tilde{\boldsymbol{a}}_j = \delta_{i^{} j}$. The magnetic
field is $\boldsymbol{B}= \ell \Phi_0 \left( \tilde{\boldsymbol{a}}_x \times
\tilde{\boldsymbol{a}}_y \right)$ with $\ell$ integer. The usual
linear{\emdash}but aperiodic{\emdash}gauge for this field would be
$\boldsymbol{A}_{\operatorname{lin}} \left( \boldsymbol{r} \right) = \ell \Phi_0 \left(
\boldsymbol{r} \cdot \tilde{\boldsymbol{a}}_x \right) \tilde{\boldsymbol{a}}_y$. A
periodic gauge can now be defined as:
\begin{eqnarray*}
  \boldsymbol{A} \left( \boldsymbol{r} \right) & = & \ell \Phi_0 \left\llbracket
  \boldsymbol{r} \cdot \tilde{\boldsymbol{a}}_x \right\rrbracket \left(
  \tilde{\boldsymbol{a}}_y - \delta \left( \left\llbracket \boldsymbol{r} \cdot
  \tilde{\boldsymbol{a}}_y \right\rrbracket \right) \tilde{\boldsymbol{a}}_x
  \right)
\end{eqnarray*}
where $\left\llbracket \cdot \right\rrbracket$ denotes the fractional part of
a real number. To make sure that the phase of every link between two atoms is
well defined, the gauge field is displaced by an infinitesimal amount such
that every atom is either left or right of the divergent line.

\tmfloatsmall{\resizebox{\columnwidth}{!}{\includegraphics{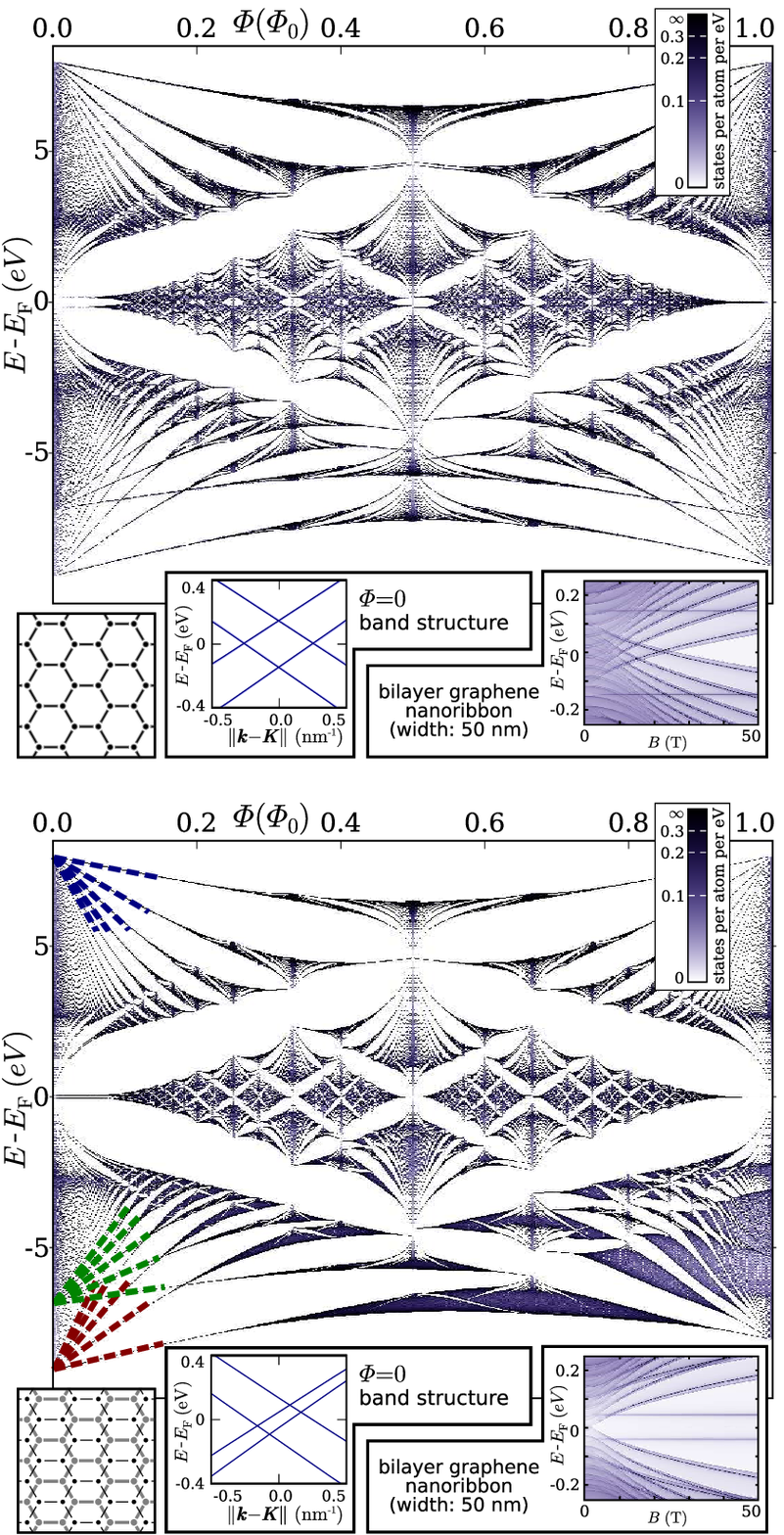}}}{\label{fig:butterflies-shifted}(Color
online) Hofstadter butterfly of a bilayer graphene in two differently shifted
configurations. Top panel: \textit{AA} stacking (two layers exactly
aligned). The band structure for this highly symmetric stacking (same
rotational symmetry as for Bernal stacking in Fig.~\ref{fig:butterfly-bernal})
shows the single-layer cone simply split up in energy. Bottom panel:
Intermediate position between Bernal and \textit{AA} stacking. The
rotational symmetry is broken and the bands split into two cones at different
offsets from the $K$ point and different energies. The straight lines overlaid
at the energy minimum and maximum are the regular Landau levels of the massive
bands. Near $E_{\mathrm{F}}$, one can make out the parabolic traces of the
relativistic Landau levels and the horizontal lines of the SUSYLLs (see text).
Insets at the lower right of each panel: DOS of a finite-width ribbon shows
the corresponding behavior in each case.}


The Hamiltonian without magnetic field{\emdash}based on a tight-binding
parametrization originally used for multiwalled carbon
nanotubes{\cite{lambin-esocct1994,nemec-hbocnpotms2006}}{\emdash}consists of a
contribution for nearest neighbors within a layer $\langle i, j \rangle$ and
one for pairs of atoms located on different sheets $\left\langle
\hspace{-0.4ex} \left\langle i, j \right\rangle \hspace{-0.4ex}
\right\rangle$:
\begin{eqnarray*}
  \mathcal{H} & = & - \sum_{\langle i, j \rangle} \gamma_{i, j}^{\operatorname{intra}}
  c^{\dag}_i c^{\phantom{\dag}}_j - \sum_{\left\langle \hspace{-0.4ex}
  \left\langle i, j \right\rangle \hspace{-0.4ex} \right\rangle} \gamma_{i,
  j}^{\operatorname{inter}} c^{\dag}_i c^{\phantom{\dag}}_j .
\end{eqnarray*}
In absence of a magnetic field, the \textit{intralayer} hopping is fixed to
$\gamma_{i, j}^{\operatorname{intra}} = \gamma_0 = 2.66 ~ \mathrm{\operatorname{eV}}$, while
the \textit{interlayer} hopping depends on the distance only,
\begin{eqnarray*}
  \gamma_{i, j}^{\operatorname{inter}} & = & \beta \exp \left( \frac{a - \left|
  \boldsymbol{r}_i -\boldsymbol{r}_j \right|}{\delta} \right),
\end{eqnarray*}
with $\beta = \gamma_0 / 8$, $a = 3.34 ~ \text{{\AA}}$, and $\delta = 0.45 ~
\text{{\AA}}$. A cutoff is chosen as $r_{\operatorname{cutoff}} = a + 5 \delta$.
Following the Peierls substitution,{\cite{peierls-ztddvl1933}} the magnetic
field $\boldsymbol{B}$ is now implemented by multiplying a magnetic phase factor
to each link between two atoms $i$ and $j$:
\begin{eqnarray*}
  \gamma_{i, j} \left( \boldsymbol{B} \right) & = & \gamma_{i, j} \left(
  \boldsymbol{B} = 0 \right) \exp \left( \mathrm{i} \frac{2 \pi}{\Phi_0}
  \int_{\boldsymbol{r}_i}^{\boldsymbol{r}_j} \boldsymbol{A}_B \left( \boldsymbol{r}
  \right) \cdot \mathrm{d} \boldsymbol{r} \right),
\end{eqnarray*}
where the integral is computed on a straight line between the atomic positions
$\boldsymbol{r}_i$ and $\boldsymbol{r}_j$.


For the bilayer graphene, we arrive thus at a periodic Hamiltonian with a
two-dimensional unit cell containing four atoms and spanning the area of one
hexagonal graphene plaquette: $A_{\operatorname{plaquette}} = (3 \sqrt{3} / 2)
d_{\operatorname{CC}}^2$, where $d_{\operatorname{CC}} = 1.42 ~ \text{{\AA}}$ is the
intralayer distance between neighboring carbon atoms. The effect of a
perpendicular magnetic field, measured in flux per plaquette $\Phi =
A_{\operatorname{plaquette}} B$, can be calculated for commensurate values $\Phi =
\left( p / q \right) \Phi_0$ ($p$, $q$ integers) by constructing a supercell
of $q$ unit cells. The corresponding Bloch Hamiltonian $\mathcal{H} \left(
\boldsymbol{k} \right)$ is a $4 q \times 4 q$ matrix that can be diagonalized
for arbitrary values of $\boldsymbol{k}$ in the two-dimensional Brillouin zone
of area $4 \pi^2 / qA_{\operatorname{plaquette}}$.


\tmfloatsmall{\resizebox{\columnwidth}{!}{\includegraphics{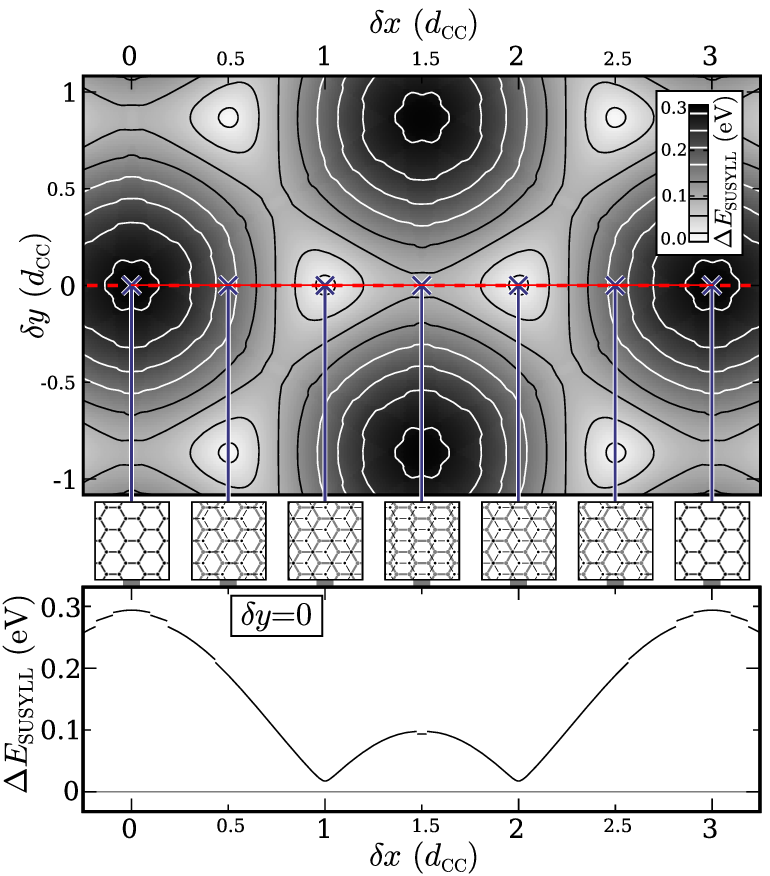}}}{\label{fig:state-splitting}(Color
online) Evolution of the split of the supersymmetric Landau level as a
function of the displacement between the two graphene layers. Top panel:
Magnitude of the split for displacements in two directions. The light spots
correspond to Bernal stacking where the level is nearly degenerate. Bottom
panel: Same data along a cut at $\delta y = 0$. The small remaining split at
the Bernal stacking configuration originates in the long-range interlayer
hoppings contained in the parametrization. The small discontinuities are
caused by the cutoff $r_{\operatorname{cut}}$. The calculation here was done at $\Phi
= \Phi_0 / 256$, but proved to be independent of the magnetic field for values
up to $\sim 0.05 \Phi_0$.}


To obtain the butterfly plots as displayed in Figs.~\ref{fig:butterfly-bernal}
and \ref{fig:butterflies-shifted}, we chose $0 \leqslant p \leqslant q = 512$,
reducing the fraction $p / q$ for efficiency. For each value of $\Phi$ the
density of states was calculated from a histogram over the spectral values for
a random sampling of $\boldsymbol{k}$ over the Brillouin zone. The number of
sampling points was chosen individually for different values of $p$ to achieve
convergence. In Figs.~\ref{fig:butterfly-bernal} and
\ref{fig:butterflies-shifted}, the Hofstadter spectra of three differently
aligned graphene bilayers are presented. The Bernal stacking
(Fig.~\ref{fig:butterfly-bernal}) stands out, as it is the configuration of
layers in natural graphite.{\cite{bernal-tsog1924,hembacher-rthaigblafm2003}}
Alternative configurations like \textit{AA}~stacking were found in
\textit{ab initio} calculations to be energetically
unfavorable;{\cite{aoki-dobsosafilg2007}} they can, however, be thought of as
either mechanically shifted samples or sections of curved bilayers
(e.g.,~sections of two shells in a large multiwall carbon nanotube) where the
alignment unavoidably varies over distance. Compared to the Hofstadter
butterfly of a single sheet of graphene,{\cite{rammal-llsobeiahl1985}} two
asymmetries are visible in all three plots: The electron-hole symmetry ($E
\leftrightarrow - E$) is broken down by the interlayer coupling already at
zero magnetic field: while the lowest-energy states of a single graphene layer
have constant phase over all atoms and can couple efficiently into symmetric
and antisymmetric hybrid states of the bilayer system, the states at high
energies have alternating phases for neighboring atoms, so interlayer
hybridization is prohibited by the second-nearest-neighbor interlayer
coupling. For low magnetic fields, two sets of Landau levels can therefore be
observed at the bottom of the spectrum, indicating a split of the massive band
of graphene at the $\Gamma$ point ($E_{\min}^0 = - 3 \gamma_0$, $m^{\ast}_0 =
2 \hbar^2 / 3 \gamma_0 d_{\operatorname{CC}}^2$) into two bands at different energies
and with different effective masses [$E_{\min}^{\pm} \approx E_{\min}^0 \pm
1.1 ~ \mathrm{\operatorname{eV}}$, $m_{\pm}^{\ast} \approx m_0^{\ast} / (1 \mp 2.1
\beta / \gamma_0)$, independent of the relative shift of the two layers; see
the straight lines overlaid in the bottom panel of
Fig.~\ref{fig:butterflies-shifted}]. At the top of the spectrum, where the
split is prohibited, only one degenerate set of Landau levels appears, as in
single-layer graphene. The original periodic symmetry along the $B$-field axis
at one flux quantum per graphene plaquette is broken down due to the smaller
areas formed by interlayer loops. The breaking of this symmetry is comparably
small in the \textit{AA}~stacking configuration
(Fig.~\ref{fig:butterflies-shifted}, top) where loops of the full plaquette
area are dominant. In the two other configurations smaller loops are more
dominant, so the periodicity is perturbed more severely. In the intermediate
configuration (Fig.~\ref{fig:butterflies-shifted}, bottom), the fractal
patterns appear slightly smeared out for high magnetic fields, due to the
reduced symmetry of the system.

The right insets of Figs.~\ref{fig:butterfly-bernal} and
\ref{fig:butterflies-shifted} display the spectra of (200,0) bilayer graphene
nanoribbons,{\cite{nakada-esigrnseaesd1996}} each in a corresponding
configuration, obtained by a method described
before{\cite{nemec-hbocnpotms2006}} that allows handling of continuous
magnetic fields.{\footnote{Adapting the conventional notation for carbon
nanotubes, an $(n$,0) ribbon has a width of $n$ hexagons and armchair edges.}}
For low magnetic fields, these spectra are strongly influenced by finite-size
effects.{\cite{wakabayashi-eamponr1999}} Only for magnetic fields larger than
$B^{\ast} \approx 4 \Phi_0 / d^2$, which for a ribbon of width $d = 50 ~
\operatorname{nm}$ relates to $\sim 7 ~ \mathrm{T}$, do the spectra of two-dimensional
bilayer graphene begin to emerge. Prominent in all three insets are the dark,
horizontal pairs of lines at the center, the supersymmetric Landau levels.
While these represent discrete levels in two-dimensional graphene sheets, they
are broadened by the finite width of the ribbon to a peak of the same shape as
in carbon nanotubes.{\cite{lee-sicniatmf2003,nemec-hbocnpotms2006}} The
mesoscopic character of these split SUSYLLs in dependence on the width $W$ of
the ribbon is captured by the functional form of the density of states:
\begin{eqnarray*}
  \rho \left( E, B, W) \right. & = & f \left( \left( E - E_0 \right) W, BW^2
  \right)
\end{eqnarray*}
where $E_0$ is the position of the maximum.


Single-layer graphene is known to feature an anomalous supersymmetric Landau
level at the Fermi
energy.{\cite{mcclure-dog1956,gusynin-uiqheig2005,ezawa-sauqheig2006}}
Neglecting Zeeman splitting, this level is fourfold degenerate (twice spin,
twice valley) and half filled. For bilayer graphene in Bernal stacking
(Fig.~\ref{fig:butterfly-bernal}) the SUSYLLs of the two layers have been
shown to be protected by symmetry and to remain degenerate, giving in total an
eightfold degeneracy.{\cite{mccann-ldaqheiagb2006}} In
Fig.~\ref{fig:butterflies-shifted}, this degeneracy can be observed to be
lifted for displaced bilayers, leading to a split of the SUSYLL into a bonding
and an antibonding hybrid state in the two layers, each fourfold degenerate.
The continuous evolution of the split for varying displacement of the two
layers against each other is displayed in Fig.~\ref{fig:state-splitting}. The
split reaches its maximum of $\Delta E \sim 0.3 ~ \mathrm{\operatorname{eV}}$ for the
\textit{AA}-stacking configuration and is minimal for Bernal stacking. For
simpler tight-binding parametrizations that take into account only first- and
second-nearest-neighbor interlayer hoppings, the degeneracy in the Bernal
configuration is known to be exact.{\cite{mccann-ldaqheiagb2006}} Here, in
contrast, this degeneracy is split by $\Delta E \sim 0.01 ~
\mathrm{\operatorname{eV}}$ due to interlayer hoppings of a longer range, similar to
the effect caused by second-nearest-neighbor interactions within one
layer.{\cite{mccann-agitebsobg2006}}


In conclusion, we have developed a method that allows the nonperturbative
implementation of a magnetic field in periodic systems with arbitrarily
positioned atoms. A $\pi$~orbital parametrization for graphitic interlayer
interactions with arbitrary displacements was then used to calculate the
Hofstadter spectrum of bilayer graphene in various configurations, revealing
common features like electron-hole symmetry breaking, and differences,
especially in the breaking of the magnetic-field periodicity. A close look at
the supersymmetric Landau level at low fields near the Fermi energy revealed a
breaking of the previously found symmetry, resulting in a split of the level,
depending on the lateral displacement of the two graphene layers against each
other.


We acknowledge fruitful discussions with I. Adagideli, C. Berger, V. Fal'ko,
F. Guinea and H. Schomerus. This work was funded by the Volkswagen Foundation
under Grant No.~I/78~340 and by the European Union program CARDEQ under
Contract No.~IST-021285-2. Support from the Vielberth Foundation is also
gratefully acknowledged.


\end{document}